\documentclass[conference, a4paper]{IEEEtran}
\IEEEoverridecommandlockouts
\usepackage{url}

	\usepackage[turkish]{babel}
	\usepackage[utf8]{inputenc} % Kullanılan encodinge göre utf8 yerine latin5 de yazılabilir.
	\usepackage[T1]{fontenc}
	
	\usepackage{float}
	
	\usepackage[skip=10pt,font=scriptsize]{caption}
	\usepackage{graphicx}
	\usepackage{threeparttable}
\usepackage{cite}
\usepackage[cmex10]{amsmath}
\usepackage{flushend}

\usepackage{xspace}
\usepackage{algorithm}
\usepackage[noend]{algpseudocode}
\floatname{algorithm}{Algoritma}

\renewcommand{\algorithmicrequire}{\textbf{Girdi:}}
\renewcommand{\algorithmicensure}{\textbf{\c{C}{\i}kt{\i}:}}

\usepackage{multirow}
\usepackage{array}
\usepackage[lofdepth,lotdepth]{subfig}

\AtBeginDocument{%
  \renewcommand\tablename{TABLO}
}

\AtBeginDocument{%
  \renewcommand\abstractname{Abstract}
}
\newcommand{\nbor}{{\tt nbor}}

\begin{document}

\IEEEpubid{\makebox[\columnwidth]{978-1-7281-7206-4/20/\$31.00 ©2020 IEEE\hfill}
\hspace{\columnsep}\makebox[\columnwidth]{}}

\title{Çizge Renklendirme Problemi için Nokta Sıralama Algoritmaları\\
Vertex Ordering Algorithms for Graph Coloring Problem}

\author{\IEEEauthorblockN{
Arda Aşık, İbrahim Buğra Demir, Berker Demirel, Barış Batuhan Topal,  Kamer Kaya*\thanks{*Equal contribution of all authors / Tüm yazarlar eşit derecede katkı sağlamıştır.}}
\IEEEauthorblockA{Sabancı Üniversitesi - İstanbul, Türkiye \\ \{ardaasik, idemir, berkerdemirel, barisbatuhan, kaya\}@sabanciuniv.edu}
}



\maketitle

\begin{ozet}
Çizge renklendirme, pratikte birçok uygulaması olan kombinatoryal ve temel bir problemdir. Bu problemde verilen bir çizge içerisindeki noktaların, her bir nokta komşuları ile farklı renkte olacak şekilde ve {\em en az renk kullanılarak} renklendirilmesi gerekmektedir. Bu optimizasyon probleminin NP-Zor olduğu kanıtlanmıştır. Bu nedenle literatürde en az sayıda renk kullanmasa da, az sayıda renk kullanmayı hedefleyen açgözlü algoritmalar bulunmaktadır. Bu algoritmalar çizgedeki noktaları belli bir sıra ile gezerek teker teker renklendirmektedir. Noktaların sırası kullanılan renk sayısı üzerinde önemli bir etkiye sahiptir. Bu çalışmada, sosyal ağ analizi metriklerinin noktalarının sırası bulunurken kullanılması üzerine yoğunlaşılmıştır. Yapılan deneyler göstermiştir ki, bir sosyal ağ analizi metriği olan {\em yakınlık merkeziyeti}, açgözlü renklendirme algoritmaları için bir sıralama ölçütü olarak kullanıldığında, algoritmalar klasik sıralama yöntemlerine göre daha az renk kullanmaktadır.
%\boldmath
\end{ozet}
\begin{IEEEanahtar}
Çizge renklendirme, açgözlü algoritmalar, yakınlık merkeziyeti.
\end{IEEEanahtar}

\begin{abstract}
Graph coloring is a fundamental problem in combinatorics with many applications in practice. In this problem, the vertices in a given graph must be colored by {\em using the least number of colors} in such a way that a vertex has a different color than its neighbors. The problem, as well as its different variants, have been proven to be NP-Hard. Therefore, there are greedy algorithms in the literature aiming to use a small number of colors. These algorithms traverse the vertices and color them one by one. The vertex visit order has a significant impact on the number of colors used. In this work, we investigated if social network analytics metrics can be used to find this order. Our experiments showed that when {\em closeness centrality} is used to find vertex visit order, a smaller number of colors is used by the greedy algorithms. 

\end{abstract}



% \fi
\IEEEpeerreviewmaketitle

\IEEEpubidadjcol

\section{G{\footnotesize İ}r{\footnotesize İ}ş}

Çizge renklendirme probleminde bir çizgenin noktaları, her bir nokta komşularından farklı renkte olacak şekilde ve en az renk kullanılarak renklendirilmelidir. Komşuluk ilişkisi tek bir kenar üzerinden tanımlandığından problemin bu hali {\em 1-uzaklık çizge renklendirme} olarak da adlandırılır. Problemin bu basit halinin bile NP-Zor olduğu kanıtlanmıştır~\cite{matula_SL,Zuckerman}. Literatürde daha genel renklendirme problemleri de incelenmiştir. Örneğin komşuluk ilişkisi $\ell$ kenar ile tanımlandığında, problem {\em $\ell$-uzaklık çizge renklendirme} problemi olarak adlandırılır. Bu problemde, iki nokta arasında $\ell$ ya da daha az kenar kullanan bir yol varsa, bu iki nokta farklı renklerde olmalıdır. 

Çizgeler gerçek hayatta birçok farklı veriyi ve problemi modellemek için kullanılmaktadır. Bu nedenle, çizge renklendirme problemlerinin pratikte birçok uygulaması bulunmaktadır. Örneğin, kablosuz bir ağdaki cihazlar çizge üzerindeki noktalar, cihazlar arasındaki potansiyel parazitler çizgede birer kenar olarak modellenebilir. Bu ağ üzerindeki kanal atama probleminde, cihazlar kanallara birbirleri ile parazit oluşturan (ya da çizge üzerinde kenar paylaşan) iki cihaz farklı kanallara atanacak şekilde yerleştirilmelidir. Bu atama yapılırken bütün cihazların (noktaların) kapsanması ve kanal (renk) sayısının en az olması istenmektedir. Bu da bir çizge renklendirme problemidir~\cite{Balasundaram06graphdomination}. Çizge renklendirme iletişim ağları üzerinde parazit azaltma, içerik iletimi, önbellek yapısının ayarlanması gibi problemler için de kullanılmıştır~\cite{hassan11, 8080217,DBLP:journals/corr/abs-1801-00106}. Bunun yanında nükleik asit dizisi tasarımı~\cite{7498232}, hava trafik akış yönetimi~\cite{Barnier2004}, sosyal ağlarda topluluk tespiti~\cite{6921552} ve paralel hesaplama~\cite{10.1145/2513109.2513110} alanlarında problemin farklı türleri karşımıza çıkmaktadır. 

Renklendirme problemi NP-Zor olduğundan literatürde bu problem için açgözlü algoritmalar önerilmiştir. 
Bu algoritmalar çizge üzerindeki noktaları belli bir sırada gezer ve her noktayı o an için komşuluğunda kullanılmayan bir renge boyar. Bu çalışmada kullanılan algoritma, noktayı kullanılmayan ilk renge boyamaktadır. 
Açgözlü algoritmaların kullandığı nokta sırası oldukça önemlidir. 
Örneğin, verilen bir çizge için elimizde en az sayıda rengi kullanan bir renklendirmenin olduğunu varsayalım. 
Noktaları renklerine göre sıralayıp (ve renkleri silip) açgözlü algoritmaya verdiğimizde algoritma bize en iyi sonucu verecektir. Fakat kötü bir sıralama, renk sayısını arttıracaktır. 
Bu makalede sosyal ağ analizinde kullanılan metriklerin, açgözlü renklendirme algoritmalarının nokta sıralamasının bulunması üzerine çalışılmıştır.
Yapılan deneyler göstermiştir ki, bir sosyal ağ metriği olan {\em yakınlık merkeziyeti} ile noktalar sıralandığında, açgözlü algoritmanın kullandığı renk sayısı azalmaktadır. 

\section{Yöntem}\label{sec:yontem}
$G = (V, E)$ yönsüz bir çizge olsun ve $V$ kümesindeki bütün $v$ noktaları için $\nbor$($v$) $\subset V$ fonksiyonu $v$'nin komşuluk kümesini belirtiyor olsun. Algoritma anlatılırken renkler, doğal sayılar ile gösterilecektir ve sayı {\bf -1} olduğunda nokta henüz renklendirilmemiş olarak düşünülecektir. Açgözlü algoritmanın sözde kodu Algoritma~\ref{alg:color}'de gösterilmiştir. Algoritma $\nbor$ fonksiyonunun tanımına göre farklı türdeki renklendirme problemleri için kullanılabilir. Komşuluk fonksiyonu $$\nbor(v) = \{u : \{v,u\} \in E\}$$ olarak tanımlandığında algoritma 1-uzaklık renklendirme problemi için bir renklendirme üretecektir. 
%Eğer $$\nbor_2(v) = \nbor_1(v) \cup  \bigcup_{u \in \nbor_1(v)} \nbor_1(u)$$ tanımlanırsa, algoritma 2-uzaklık renklendirme problemi için kullanılabilir. 

Komşuluk fonksiyonundan bağımsız olarak, açgözlü algoritma ilk noktalarda renk seçimi açısından daha rahat olacak, fakat noktalar boyandıkça renk seçiminde zorlanmaya başlayacaktır. Dolayısıyla noktaların sıralaması oldukça önemlidir. Rastgele bir sıralama her zaman iyi sonuç vermeyebilir. Literatürde bu algoritmanın sıralama özelinde analizi yapılmış, rastgele sıralamaların yüksek olasılıkla minimum renk sayısından çok daha fazla renk kullandığı çizge çeşitlerinin varlığı gösterilmiştir~\cite{KUCERA1991674,DBLP:journals/corr/Husfeldt15}.  
 
\begin{algorithm}
\small
\caption{\textsc{ÇizgeRenklendirme}}
\algorithmicrequire{ $G = (V, E)$, $V$: renklendirilecek noktalar, $E$: kenarlar\\\hspace*{7ex}$\nbor$($.$): komşuluk fonksiyonu}\\
\algorithmicensure{ $c[.]$:  renklendirme listesi.}
\begin{algorithmic}[1]
\For{her $v \in V$ noktası için} 
\State{$F \leftarrow \emptyset$} 
\For{her $u \in$ $\nbor$($v$)}
\If{$c[u] \neq {\mathbf {-1}}$}
\State{$F \leftarrow F \cup \{c[u]\}$}
\EndIf
\EndFor
\State{$col \leftarrow 0$} \Comment{ilk uygun renk yöntemi}
\While{$col \in F$}
\State{$col \leftarrow col + 1$}
\EndWhile
\State{$c[v] \leftarrow col$}
\EndFor
\end{algorithmic}
\label{alg:color}
\end{algorithm}

Literatürde noktaların sıralanması için farklı yöntemler denenmiştir. Sık kullanılan bir yöntem,
noktaların bağlı oldukları kenar sayıları (dereceleri) kullanılarak sıralanmasıdır. 
Noktalar azalan kenar sayısına göre 
sıralandığında daha riskli noktalar önce boyanacak ve algoritma daha kısıtlanmış olduğu son adımlarda 
renk sayısını arttırmadan daha bir çözüm elde edebilecektir. Bu sıralamaya Welsch-Powell yöntemi de denmektedir~\cite{wwws}. 
Noktalar derecelerine, $|\nbor_1(.)|$ değerlerine göre sıralandıklarında açgözlü algoritmanın kullanacağı maksimum renk sayısı ${{\tt max}_{v \in V}}\{d_v\} + 1$ olacaktır. 

Bu çalışmada kullanılan sıralama yöntemleri aşağıda verilmiştir.

\begin{itemize}
    \item {\bf 1-komşuluk}: Bu yöntem yukarıda anlatılan, azalan $|\nbor(.)|$ değerlerini kullanan sıralama yöntemidir. 
   
    \item {\bf 2-komşuluk}: Bu yöntem noktaları azalan $|\nbor_2(.)|$ değerlerine göre sıralar. Çizge üzerindeki herhangi iki $v$ ve $u$ noktası arasındaki en kısa uzaklık $d(v,u)$ ile gösterilsin. Bir $v \in V$ noktasının 2-komşuluğu, $$\nbor_2(v) = \{u \in V: d(v,u) = 2\}$$ olarak hesaplanır.
   
    \item {\bf 3-komşuluk}: Bu yöntem noktaları azalan $|\nbor_3(.)|$ değerlerine göre sıralar. Bir $v \in V$ noktası için $\nbor_3(v)$ kümesi $$\nbor_3(v) = \{u \in V: d(v,u) = 3\}$$ olarak hesaplanır.
   
    \item {\bf Yakınlık merkeziyeti}: Çizge üzerindeki herhangi iki $v$ ve $u$ noktası arasındaki en kısa uzaklık $d(v,u)$ ile gösterilsin. Bir $v$ noktasının yakınlık merkeziyeti $${\tt ym}(v) = \frac{1}{\sum_{u \in V} d(v,u)}$$ olarak hesaplanır~\cite{Sabidussi1966,doi:10.1121/1.1906679}. Bu yöntem noktaları azalan yakınlık merkeziyeti değerlerine göre sıralar. Yöntemin amacı çizgeyi {\em içeriden dışarıya} doğru, öncelikle içerideki yoğun kısmın üzerinden geçerek renklendirmektir. Dolayısıyla çizgedeki zor ve riskli noktalar daha önce renklendirilecek, dışarıda kalan, kullanılan renk sayısını arttırması daha düşük olasılığa sahip uç noktalar algoritmanın kısıtlandığı son adımlarda renklendirilecektir. 
   
     \item {\bf Kümeleme katsayısı}: Bu yöntemde noktalar azalan {\em kümeleme katsayısı} değerlerine göre sıralanır. Kümeleme katsayısı, noktanın komşularının birbirine ne kadar bağlı olduğunu gösterir~\cite{Watts1998Collective,doi:10.1177/104649647100200201}. Çizge üzerindeki bir $v$ noktası için $${\tt kk}(v) = \frac{|\{\{u,w\} \in E: u,w \in \nbor(v)\}| }{|\nbor(v)| (|\nbor(v)| - 1)}$$ noktanın kümeleme katsayısını vermektedir. Bu değer yüksek olduğunda nokta neredeyse birbirine tam bağlı bir alt çizge içerisinde yer almaktadır. Bu da noktayı açgözlü algoritma için riskli bir nokta yapmaktadır. Bu sıralama yöntemi bu tür noktaları ilk sıralara koyarak renk sayısını azaltmayı hedeflemektedir. 
   
     \item {\bf PageRank}: PageRank sosyal ağ analizinde sıklıkla kullanılan bir merkeziyet metriğidir~\cite{Pageetal98}. Çizge üzerindeki bir noktanın PageRank değeri, kenarlar üzerinden rastgele nokta gezintisi yapan bir kişinin belli bir anda o noktada bulunma olasılığını vermektedir.
     Bu sıralama yöntemi azalan PageRank değerlerini kullanır. Bu yöntemin temel motivasyonu, yakınlık merkeziyeti için kullanılan yöntemin motivasyonu ile aynıdır ve çizgenin merkezinden başlayarak, renk sayısını arttırma olasılığı yüksek olan noktaları ilk sırada renklendirmektir. Çizge üzerindeki noktaların PageRank değerleri yinelemeli bir algoritma ile hesaplanır. Bir $v$ noktasının başlangıçtaki PageRank degeri ${\tt pr}_0(v) = \frac{1}{|V|}$ olarak kabul edilir. Hesaplama esnasında $i$ adımındaki PageRank değeri ise
$${\tt pr}_i(v) = \frac{1-\alpha}{|V|} + \alpha \sum_{u \in \nbor_1(v)} \frac{{\tt pr}_{i-1}(u)}{|\nbor_1(u)|}
$$ olur. 

     \item {\bf Rastgele}: Bu yöntem noktaları rastgele, gelişigüzel bir şekilde sıralamaktadır. 
\end{itemize}

\section{Deney Sonuçları}
Bu çalışmada yapılan deneyler, 512GB RAM'e sahip, 64 bit CentOS 6.5 ile çalışan, her bir soketin 15 çekirdeğe sahip olduğu (toplamda 60), 2.30 GHz ile çalışan Intel Xeon E7-4870 v2'de yapıldı. Yazılan bütün kodlar {\tt gcc 8.2.0} ile optimizasyon parametresi {\tt -O3} kullanılarak derlendi.

Deneylerde kullanılan bütün çizgeler SuiteSparse\footnote{\url{http://faculty.cse.tamu.edu/davis/suitesparse.html}} seyrek matris kütüphanesinden indirildi. İlk kısım deneyler için nokta sayısı $10^5$ ila $10^6$ arasındaki bütün simetrik matrisler kullanıldı. Deneyler yapılırken, kütüphanede bu özelliklere sahip 204 matris bulunmaktaydı. Her bir matris, noktalar matrisin satır ve sütunları, kenarlar da matris içerisindeki sıfırdan farklı sayılar olmak üzere bir çizge şeklinde modellendi. Bir sıralama yönteminin bir matris özelindeki performansı ölçülürken, o yöntem  ile elde edilen renk sayısının, 1-komşuluk yöntemi ile elde edilen renk sayısına oranı kullanıldı. Performans sonuçları ölçülürken, PageRank için 20 yineleme yapılmıştır. Rastgele sıralama yönteminin performansı ise 5 rastgele permütasyonun ortalaması alınarak ölçülmüştür. Yakınlık merkeziyeti için ise sonucu kesin olmasa da çok hızlı şekilde veren bir yakınsama algoritması kullanılmıştır~\cite{doi:10.1142/S0218127407018403}. 

\begin{figure}[H]
    \centering
      \shorthandoff{=}
    \includegraphics[width=0.95\linewidth]{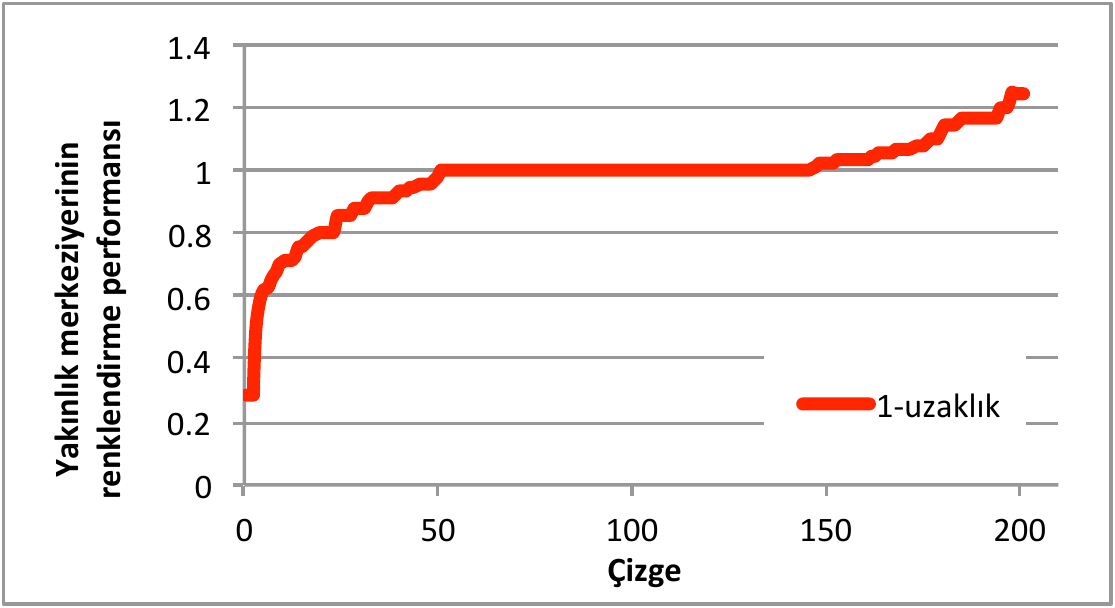}
    \caption{Yakınlık merkeziyeti metriğinin renklendirme problemi için 204 çizge üzerindeki performansı.}
    \label{fig:d1d2}
\end{figure}

Tablo~\ref{tab:big_graphs}'de, Bölüm~\ref{sec:yontem}'de anlatılan sıralama yöntemlerinin bütün matrisler için hesaplanan performanslarının geometrik ortalamaları verilmiştir. Tabloda da görüldüğü üzere, Yakınlık Merkeziyeti tabanlı sıralama diğer tüm sıralamalardan daha etkili bir sonuç vermiştir. Bu sonuçların daha detaylı incelemesi için Şekil~\ref{fig:d1d2}'e bakılabilir. Görüldüğü gibi, yakınlık merkeziyeti tabanlı sıralama kimi çizgeler için 4.5 kat daha az renk kullanırken, bu çizgeler üzerinde en fazla $\%25$ fazla renk kullanmaktadır. 
\vspace*{-2ex}
\begin{table}[H]
    \centering
    \fontsize{9}{9}\selectfont
    \caption{1-UZAKLIK RENKLENDİRME İÇİN SIRALAMA ALGORİTMALARININ PERFORMANSLARININ GEOMETRİK ORTALAMASI}
    \def\arraystretch{1.3}
    \scalebox{0.99}{
    \begin{tabular}{l|r}
        %\hline
        {\bf Sıralama} & {\bf1-uzaklık} \\
        \hline
        1-komşuluk & 1.00 \\ 
        2-komşuluk & 1.02 \\ 
        3-komşuluk & 1.07 \\ 
        Yakınlık Merkeziyeti & {\bf 0.97} \\ 
        Kümelenme Katsayısı & 1.03 \\ 
        PageRank & 1.02  \\ 
        Rastgele Sıralama & 1.08 \\
 %       \hline
    \end{tabular}
    }
    \label{tab:big_graphs}
\end{table}

Her ne kadar yakınlık merkeziyeti tabanlı sıralama bütün sıralama yöntemlerinden daha iyi çalışmış olsa da, performans sadece 1-komşuluk sıralama yöntemine göre değerlendirilmiştir. Bunun sebebi, kullanılan çizgelerin en iyi çözümü hızlı bir şekilde elde edemeyeceğimiz kadar büyük olmasıdır. Performansın en iyi çözüme göre değerlendirilmesi için ikinci, görece daha küçük çizgelerden oluşan bir çizge kümesi oluşturulmuştur. Bu kümeyi oluşturmak için SuiteSparse kütüphanesinden nokta sayısı 100 ila 500 arası, 260 simetrik matris seçilmiş ve çizge olarak modellenmiştir. Bütün bu çizgeler için en az sayıda renk kullanan çözümler, lineer programlama yöntemi ve {\em CPLEX} kütüphanesi kullanarak elde edilmiştir. 

Bu küme üzerindeki performans sonuçları Tablo~\ref{tab:small_graphs}'de verilmiştir. Değerlerden görüleceği üzere, yakınlık merkeziyeti büyük çizgelerde olduğu gibi daha önce anlatılan bütün sıralama yöntemlerinden daha başarılı bir performans elde etmiştir. Yakınlık merkeziyeti, en iyi sonuca göre ortalama $\%23$ daha fazla renk kullanmaktadır. 1-komşuluk tabanlı sıralama ise en iyi sonuca göre $\%36$ daha fazla renkle çizge  renklendirme yapabilmektedir. 

\begin{table}[H]
    \centering
    \fontsize{9}{9}\selectfont
    \caption{1-UZAKLIK RENKLENDİRME İÇİN SIRALAMA ALGORİTMALARININ EN İYİ ÇÖZÜME GÖRE PERFORMANSLARININ GEOMETRİK ORTALAMASI}
    \def\arraystretch{1.3}
    \scalebox{0.99}{
    \begin{tabular}{l|r}
        % \hline
        {\bf Ölçüt} & {\bf1-uzaklık} \\
        \hline
        1-komşuluk & 1.36  \\ 
        2-komşuluk & 1.35 \\ 
        3-komşuluk & 1.40 \\ 
        Yakınlık Merkeziyeti & {\bf 1.23} \\ 
        Kümelenme Katsayısı & 1.40 \\
        PageRank & 1.34 \\ 
        Rastgele Sıralama & 1.56 \\
        \hline
        Ağırlıklı Sıralama & {\bf 1.16} \\ 
        Tekdüze Sıralama & 1.53 \\
        % \hline
    \end{tabular}}
    \label{tab:small_graphs}
\end{table}
    
Bu küme üzerindeki performans sonuçlarının oluşturulması sürecinde büyük çizgelerde kullanılan sıralama algoritmalarına ek olarak {\em tekdüze} ve {\em ağırlıklı} sıralama yöntemleri de denenemiştir. Bu sıralama yöntemleri normalize edilmiş değerleri kullanır. Bunun için bütün nokta değerlerinden o değerin ortalaması çıkartılır ve elde edilen değer standart sapmaya bölünür. {\em Tekdüze sıralama} bir $v$ noktasının normalize edilmiş bütün değerlerini~(1-komşuluk, 2-komşuluk, 3-komşuluk, yakınlık merkeziyeti, kümelenme katsayısı ve PageRank) toplayarak elde edilen değer için azalan sıraya göre noktaları sıralar. Tablo~\ref{tab:small_graphs}'de görüldüğü gibi bu sıralama rastgele bir sıralama kadar kötü sonuçlar vermiştir.

{\em Ağırlıklı sıralama}, Tablo~\ref{tab:small_graphs}'de verilen metriklerin farklı kombinasyonlarının kullanılması ile en fazla ne kadar iyi bir sonuç alınabileceğini ölçmek için tasarlanmıştır. Sıralamalar, birbirleri arasındaki ölçek farkını gidermek için normalize edildikten sonra, tüm lineer kombinasyonlar arasından en iyi ortalama sonucu veren ağırlık kombinasyonu bulunmuştur. Elde edilen ağırlıklar Tablo~\ref{tab:weights}'te verilmiştir.  

\begin{table}[H]
    \centering
    \fontsize{9}{9}\selectfont
    \caption{AĞIRLIKLI SIRALAMA ALGORİTMASI İÇİN AĞIRLIK DEĞERLERİ}
    \def\arraystretch{1.3}
    \scalebox{0.99}{
    \begin{tabular}{l|r}
        % \hline
        {\bf Ölçüt} & {\bf1-uzaklık} \\
        \hline
        1-komşuluk & 0.10  \\ 
        2-komşuluk & 0.05 \\ 
        3-komşuluk & 0.10 \\ 
        Yakınlık Merkeziyeti & {\bf 0.70} \\ 
        Kümelenme Katsayısı & 0.05 \\ 
        PageRank & 0.00 \\
    \end{tabular}}
    \label{tab:weights}
\end{table}

Katsayılar incelendiğinde, yakınlık merkeziyeti değerinin diğer ağırlıklardan daha yüksek olduğu görülür. Başka bir deyişle, yakınlık merkeziyeti, en iyi kombinasyon içerisinde diğer metriklerden daha büyük bir etkiye sahiptir. Bu sıralama algoritması ile en iyi sonuçtan sadece $\%16$ kötü sonuç elde edilmiştir. Yine Tablo~\ref{tab:small_graphs}'de görüldüğü gibi, yakınlık merkeziyeti, küçük çizgeler için sıralama algoritmalarıyla oluşturulabilecek en iyi renklendirmeden ortalama sadece $\%7$ daha fazla renk ile çizgeleri renklendirebilmektedir.

\begin{figure}[H]
    \centering
      \shorthandoff{=}
    \includegraphics[width=0.95\linewidth]{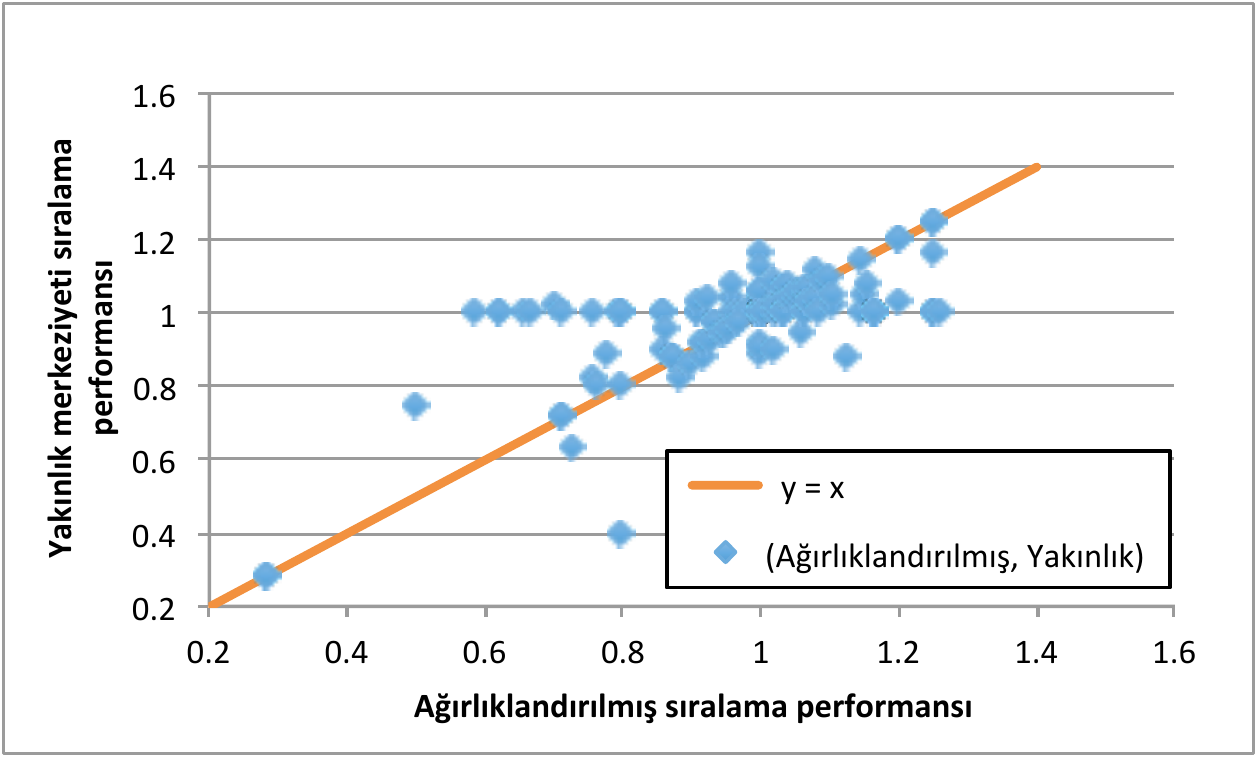}
    \caption{Ağırlıklandırılmış sıralamanın ve yakınlık merkeziyeti metriğinin renklendirme problemi için 204 çizge üzerindeki performansı.}
    \label{fig:yw}
\end{figure}

Ağırlıklandırılmış sıralama ilk deneydeki çizge kümesi üzerinde kullanıldığında, yakınlık merkeziyeti tabanlı sıralamanın iyileştiremediği çok sayıda çizgeyi iyileştirdiği görülmüştür (Şekil~\ref{fig:yw}). Ağırlıklandırılmış sıralama yöntemi, 204 çizgenin sadece 45'inde yakınlık merkeziyeti tabanlı sıralamadan daha iyi sonuç elde edememiştir. Dolayısıyla, ağırlıklandırılmış yöntemin başarılı bir sıralama yaratma potansiyelinin olduğu düşünülmektedir.  

\section{Sonuç}

Bu çalışmada çizge renklendirme problemi için kullanılan açgözlü algoritmalardaki nokta sıralaması için yöntemler üzerine yoğunlaşılmıştır. Sosyal ağ analizi için kullanılan metriklerin bu sıralamaların bulunmasındaki başarıları araştırılmıştır. Yapılan deneyler göstermiştir ki, yakınlık merkeziyeti metriği, sıralama için oldukça başarılı bir metriktir. 

Bu çalışmada sadece statik sıralama yöntemleri üzerinde durulmuştur. Algoritma çalıştıkça noktalar renklendirildiğinden, her adımda farklı bir sıralamanın en iyi olma olasılığı oldukça fazladır. İleride, sıralamanın noktaların komşuluklarındaki farklı renk sayısı gibi dinamik metrikler de kullanılrarak, sürekli değişen sıralamalar üzerinde çalışılacaktır. Verilen bir çizge için en iyi sıralama yönteminin tahmin edilmesi de oldukça ilginç bir problemdir. İleride bu problem üzerinde de çalışılması planlanmaktadır.

\bibliography{bare_conf}
\bibliographystyle{plain}
\end{document}